\documentclass[a4paper]{jpconf}
\usepackage{graphicx}
\usepackage{xspace}
\usepackage{amsmath}
\usepackage{amssymb}
\usepackage{tikz}
\usetikzlibrary{matrix,shapes,arrows,positioning,chains}

\newcommand{\fbinv} {\mbox{\ensuremath{\,\text{fb}^{-1}}}\xspace}
\newcommand{\GeV}{\ensuremath{\,\mathrm{Ge\hspace{-.08em}V}}\xspace}
\newcommand{\Pt}{\ensuremath{p_{\mathrm{T}}}\xspace}
\newcommand{\ttbar}{t\ensuremath{\mathrm{\bar{t}}}\xspace}

\begin{document}

\tikzset{
    block/.style={rectangle, draw, text width=21ex, text centered, rounded corners}
}

\title{Segment Linking: A Highly Parallelizable Track Reconstruction Algorithm for HL-LHC}

\author{P Chang$^{\,1}$, P Elmer$^{\,2}$, Y Gu$^{\,1}$, V Krutelyov$^{\,1}$, G Niendorf$^{\,3}$, M Reid$^{\,3}$, B V Sathia Narayanan$^{\,1}$, M  Tadel$^{\,1}$, E Vourliotis$^{\,1a}$, B Wang$^{\,2}$, P Wittich$^{\,3}$, A Yagil$^{\,1}$}

\address{$^{1\,}$University of California San Diego, CA, US\\
$^{2\,}$Princeton University, NJ, US\\
$^{3\,}$Cornell University, NY, US}

\ead{$^{a\,}$emmanouil.vourliotis@cern.ch}

\begin{abstract}
The High Luminosity upgrade of the Large Hadron Collider (HL-LHC) will produce particle collisions with up to 200 simultaneous proton-proton interactions. These unprecedented conditions will create a combinatorial complexity for charged-particle track reconstruction that demands a computational cost that is expected to surpass the projected computing budget using conventional CPUs. Motivated by this and taking into account the prevalence of heterogeneous computing in cutting-edge High Performance Computing centers, we propose an efficient, fast and highly parallelizable bottom-up approach to track reconstruction for the HL-LHC, along with an associated implementation on GPUs, in the context of the Phase 2 CMS outer tracker. Our algorithm, called Segment Linking (or Line Segment Tracking), takes advantage of localized track stub creation, combining individual stubs to progressively form higher level objects that are subject to kinematical and geometrical requirements compatible with genuine physics tracks. The local nature of the algorithm makes it ideal for parallelization under the Single Instruction, Multiple Data paradigm, as hundreds of objects can be built simultaneously. The computing and physics performance of the algorithm has been tested on an NVIDIA Tesla V100 GPU, already yielding efficiency and timing measurements that are on par with the latest, multi-CPU versions of existing CMS tracking algorithms.
\end{abstract}

\section{Introduction and Motivation} \label{sec:IntroAndMotivation}

The Large Hadron Collider (LHC) of CERN is currently the largest accelerator in the world. In order to take advantage of the extensive infrastructure for its operation for as long as possible, a major upgrade to the LHC is planned for 2027 (Phase 2). This upgrade, called the High Luminosity LHC (HL-LHC), aims at collecting more than $3000 \fbinv$ of integrated luminosity of proton-proton (p-p) collision data, increasing the corresponding nominal value of the LHC by a factor of ten~\cite{HLLHC}. To achieve such high values of integrated luminosity, the instantaneous luminosity, i.e. rate of collisions, is pushed to its limits. This leads to an increased pileup (PU), i.e. number of simultaneous p-p interactions per bunch crossing. The average PU during the full LHC operation (including Run 3) is expected to stay under 70, while the same number for the HL-LHC will rise to around 200.

The much harsher PU conditions anticipated in the HL-LHC lead to an escalation of the complexity of event reconstruction. Especially for the reconstruction of charged-particle tracks, which relies on the combination of multiple hits and is an inherently combinatorial problem, several times more PU results in unprecedented tracker occupancies and computational complexity. As shown in reference~\cite{HighPUTracking}, this translates to an exponential growth of time needed for the track reconstruction. To mitigate this effect, increased computational resources would be required, leading to highly elevated costs of operation.

Despite the projected increase in the processing power of central processing units (CPUs), which dominated the computational resources of Run 1 and 2 of the LHC, the computational complexity of event reconstruction at the HL-LHC is expected to demand a computational cost that exceeds the computing budget using conventional single-thread CPUs programming. As a result, alternative approaches need to be explored. A solution to this problem is the efficient parallelization of existing reconstruction algorithms, where this is feasible, or the creation of new algorithms that can benefit from parallelism. The concept of parallelization can be combined with the modern paradigm of heterogeneous computing, which aims at utilizing a large variety of processing units, such as graphics processing units (GPUs) or field-programmable gate arrays (FPGAs).

This paradigm shift is already implemented in Run 3 related applications in experiments of the LHC. Taking the example of the Compact Muon Solenoid (CMS) experiment~\cite{CMS}, the pixel track and vertex reconstruction (Patatrack~\cite{patatrack}), the outer tracker strip local reconstruction, and the electromagnetic calorimeter and hadronic calorimeter reconstruction algorithms have been ported to GPUs. The combination of CPU and GPU usage leads to $\sim\!25\%$ timing reductions for the total Run 3 CMS High Level Trigger (HLT) reconstruction. The HL-LHC cost projections show that offloading $50\%$~($80\%$) of the computational work to GPUs decreases the HLT processing farm cost by $35\%$~($75\%$)~\cite{CMSHLTDAQPhase2Upgrade}.

Moving in this direction, this work presents a highly parallelizable algorithm for track reconstruction at the HL-LHC, called \emph{Segment Linking} or \emph{Line Segment Tracking (LST)}. The algorithm has been developed to benefit from heterogeneous computing architectures, more specifically from GPU performance, and a relevant implementation of it on an NVIDIA Tesla V100 GPU is described.

\section{The Segment Linking Algorithm} \label{sec:TheSegmentLinkingAlgo}

The track reconstruction at the LHC is usually based on Kalman filter methods~\cite{KalmanFilter}, which are inherently sequential in nature. This has repercussions in terms of computational time for their application. For example, for the CMS experiment, track reconstruction used to require almost $60\%$ of the time needed for the whole event reconstruction procedure at 50 PU. There has been a recent effort towards the parallelization of Kalman filter approaches (\emph{mkFit} project~\cite{mkFit}) that improves the timing performance with comparable physics results. Nevertheless, there is a clear need for track reconstruction algorithms with architectures specifically designed for parallelization.

LST is an algorithm developed with the concept of parallelization built into it. The algorithm relies on local hits in the tracker to start producing short tracks. Given the local nature of these short tracks, the algorithm utilizes the Single Instruction, Multiple Data (SIMD) paradigm, where all of the short tracks can be created concurrently with the same set of instructions, since they are independent from one another. Progressively, the short tracks are linked together to form longer tracks, leading to collections of objects with different characteristics. Finally, objects from these collections are combined to create a Track Candidate collection with a high efficiency and a low fake rate. LST is inspired by the XFT algorithm in the Collider Detector at Fermilab (CDF) at the Tevatron~\cite{XFT} and its prototype was presented at the International Conference of High Energy Physics (ICHEP) in 2016~\cite{SDLFirstProc}.

\begin{figure}[tbh!]
\begin{center}
    \begin{minipage}{0.45\textwidth}
        \vspace{-5ex}
        \includegraphics[width=\textwidth]{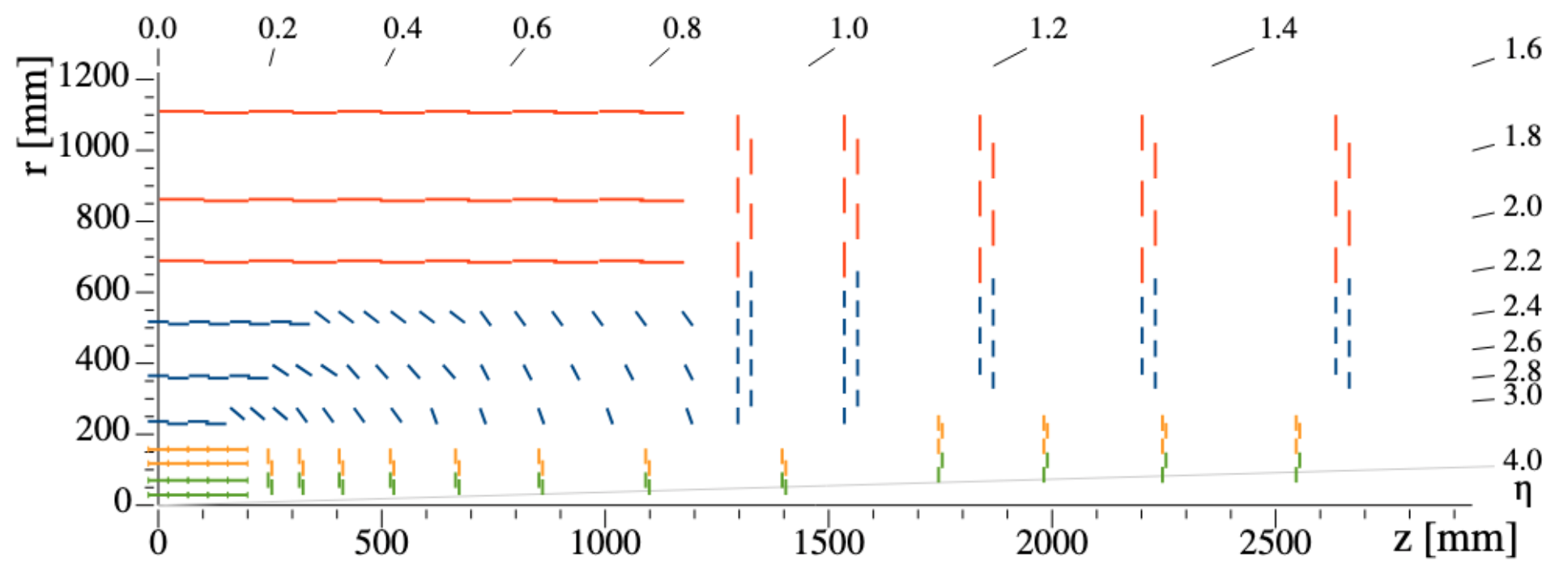}
        \caption{\label{fig:CMSTrackerPhase2}$r-z$ perspective of one quarter of the Phase-2 CMS tracker. The green and orange lines represent the inner tracker modules, while the blue and red lines represent the outer tracker modules~\cite{CMSTrackerPhase2Upgrade}.}
    \end{minipage}
    \hspace{2pc}
    \begin{minipage}{0.45\textwidth}
        \includegraphics[width=\textwidth]{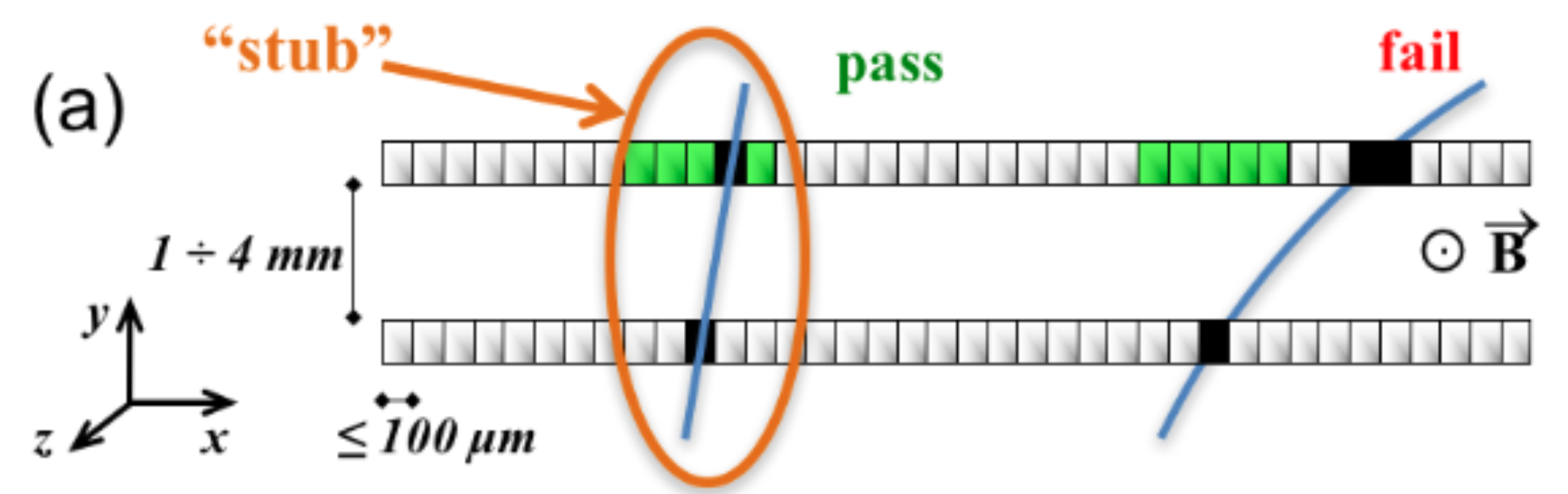}
        \caption{\label{fig:pTModule}Sketch of a tracker \Pt\ module. The series of gray squares represent the closely-spaced sensors of a module, while the green squares indicate the channels compatible with a particle which has passed through reference channel (black square). The lower the particle \Pt, the lower the probability to produce an accepted stub~\cite{CMSTrackerPhase2Upgrade}.}
    \end{minipage}
\end{center}
\end{figure}

Before delving into the details of the track building logic of the algorithm, it is worth noting a real example for its application. The CMS Phase 2 Outer Tracker, shown in figure~\ref{fig:CMSTrackerPhase2} (blue and red lines) is a suitable detector for the LST algorithm. This comes from the fact that each layer is composed by multiple ``\Pt modules" (lines in figure~\ref{fig:CMSTrackerPhase2}), each with two closely-spaced sensors which can provide a very localized combination of hits, called \emph{stub}, as shown in figure~\ref{fig:pTModule}. In the LST algorithm presented here, a stub is accepted only if the hits are within a selection window (marked in green in figure~\ref{fig:pTModule}) which corresponds to a \Pt\ threshold of $0.8 \GeV$ for the traversing particle. The usage of stubs instead of individual hits can significantly help in reducing the combinatorics in the initial steps of the LST algorithm up to 7.5 times. As an example, the number of hits in the first layer of the outer track in one \ttbar event at 200 PU event is approximately 36000, while this number goes down to 6000 for stubs in the same layer.

\vspace{-5ex}

\begin{figure}[tbh!]
    \begin{center}
        \includegraphics[width=0.45\textwidth]{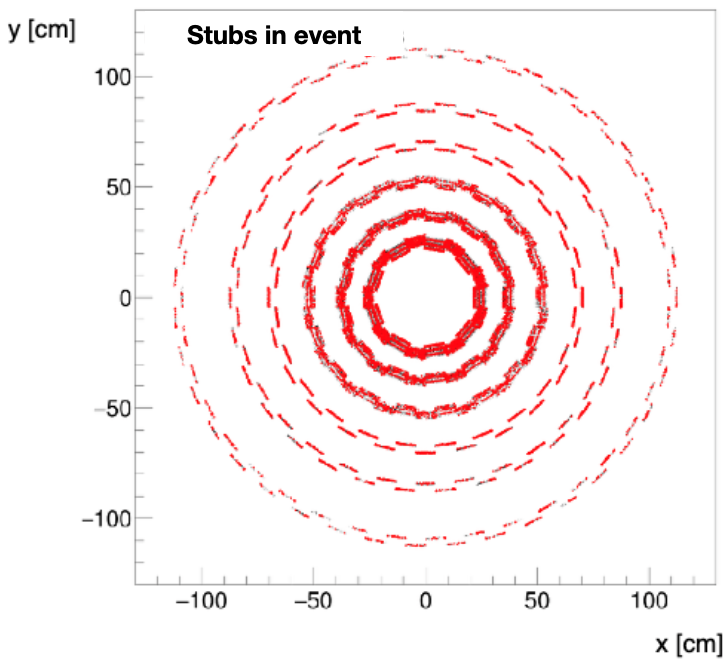}
        \hspace{2pc}
        \begin{minipage}[b]{0.45\textwidth}
            \caption{\label{fig:StubDistribution}Occupancy distribution for stubs in a single \ttbar event at 200 PU in the barrel part of the Phase-2 CMS tracker as a function of the $x$ and $y$ coordinates of the CMS detector.}
        \end{minipage}
    \end{center}
\end{figure}

\subsection{The Logic of the Algorithm} \label{sec:TheSegmentLinkingAlgo:AlgoLogic}

\begin{figure}[tbh!]
\begin{center}
    \begin{minipage}{0.45\textwidth}
        \includegraphics[width=0.95\textwidth]{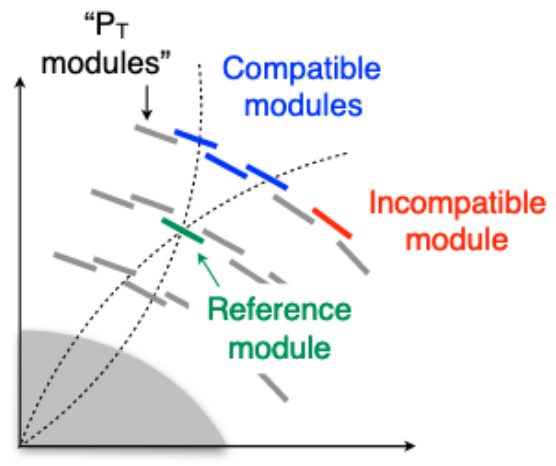}
        \caption{\label{fig:CompModules}Demonstration of the logic behind the construction of a map of modules compatible for linking algorithm objects. The dashed line represents helices of positively and negatively particles with $\Pt = 0.8 \GeV$.}
    \end{minipage}
    \hspace{2pc}
    \begin{minipage}{0.45\textwidth}
        \vspace{-3ex}
        \includegraphics[width=\textwidth]{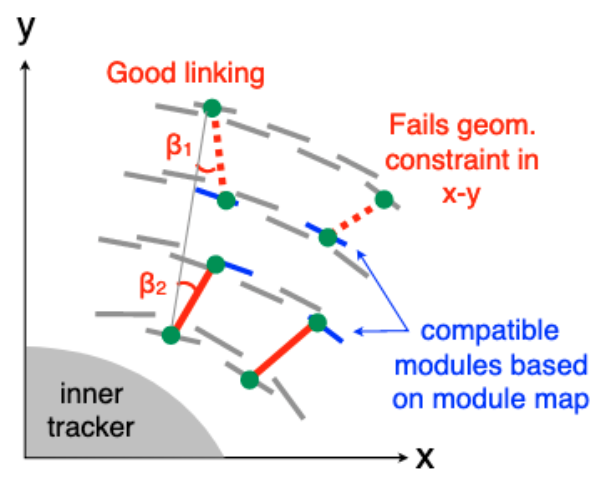}
        \vspace{-4ex}
        \caption{\label{fig:GeomReqs}Two examples showcasing how geometrical requirements, in combination with the list of compatible modules, lead to accepting or rejecting the linking of algorithm objects.}
    \end{minipage}
\end{center}
\end{figure}

\begin{figure}[tbh!]
\begin{center}
    \begin{minipage}{0.45\textwidth}
        \includegraphics[width=\textwidth]{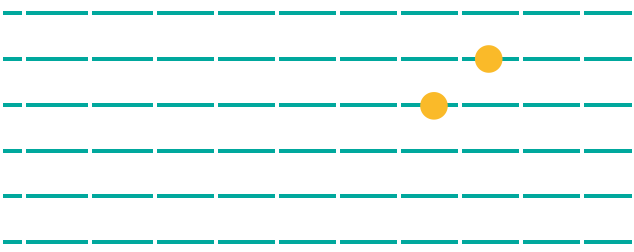}
        \caption{\label{fig:Stub}Stubs.}
    \end{minipage}
    \hspace{2pc}
    \begin{minipage}{0.45\textwidth}
        \includegraphics[width=\textwidth]{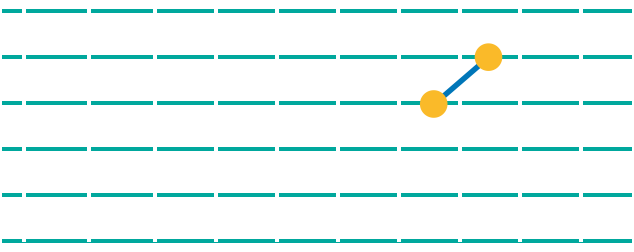}
        \caption{\label{fig:Segment}Segment.}
    \end{minipage}
    \vspace{1em}
    \hrule
    \vspace{1em}
    \begin{minipage}{0.45\textwidth}
        \includegraphics[width=\textwidth]{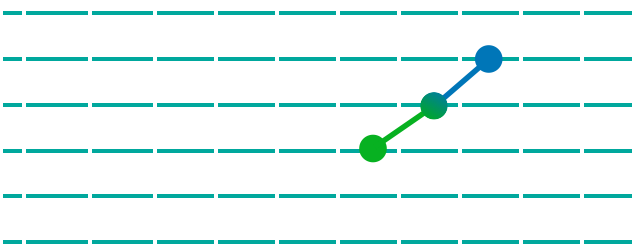}
        \caption{\label{fig:Triplet}Triplet.}
    \end{minipage}
    \hspace{2pc}
    \begin{minipage}{0.45\textwidth}
        \includegraphics[width=\textwidth]{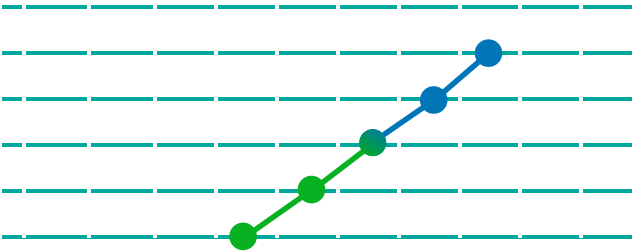}
        \caption{\label{fig:Quintuplet}Quintuplet.}
    \end{minipage}
    \vspace{1em}
    \hrule
    \vspace{1em}
    \begin{minipage}{0.45\textwidth}
        \includegraphics[width=\textwidth]{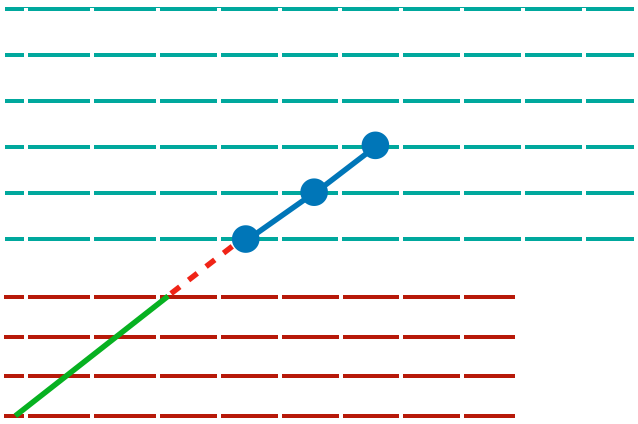}
        \caption{\label{fig:PixelTriplet}Pixel triplet.}
    \end{minipage}
    \hspace{2pc}
    \begin{minipage}{0.45\textwidth}
        \includegraphics[width=\textwidth]{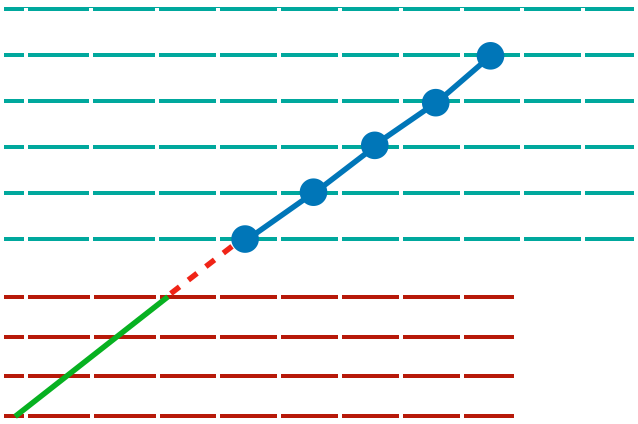}
        \caption{\label{fig:PixelQuintuplet}Pixel quintuplet.}
    \end{minipage}
\end{center}
The different objects created in the LST algorithm. The cyan lines represent the outer tracker layers, while the dark red lines represent the inner tracker layers.
\end{figure}

The track building procedure starts with the creation of stubs in each of the \Pt module of the outer tracker. The occupancy distribution of stubs in the $x-y$ two-dimensional plane of the CMS coordinate system is shown in figure~\ref{fig:StubDistribution} for a typical \ttbar event at 200 PU. The next step is the linking of stubs in neighboring layers (figure~\ref{fig:Stub}) together to form a \emph{segment} (figure~\ref{fig:Segment}). With around 6000 stubs in a single layer, the combinatorics for all the stub connections is immense. To decrease the number of combinations, physics requirements are taken into account. More specifically, helices of particles originating from the center of the detector, with both positive and negative sign hypotheses and with \Pt at the previously mentioned threshold of $0.8 \GeV$, are considered. Tracks from simulation are considered as well, in order to include cases that may have been missed by the idealistic physics scenario of helices. This leads to restrictions in the number of modules that can be linked to a single (reference) module, as it can be seen in figure~\ref{fig:CompModules}.

The creation of segments is followed by the creation of \emph{triplets (T3)} and \emph{quintuplets (T5)}. Triplets are formed from the linking of two segments that share a common stub (figure~\ref{fig:Triplet}), while quintuplets involves the linking of two triplets with a common stub (figure~\ref{fig:Quintuplet}). As objects get longer, more handles can be utilized for determining whether their linking is consistent with a track hypothesis. As an example, figure~\ref{fig:GeomReqs} shows that the relative angle between segments can be used to reject their combination if that leads to an unphysical track pattern, such as the right set of segments in the figure.

In addition to the outer-tracker-only objects, exploiting information from the inner tracker can assist in creating collections with higher purity of physical tracks. Triplets and quintuplets are linked to seeds from the pixel detector, called \emph{pixel line segments (pLS)}, hence producing even longer objects, \emph{pixel triplets (pT3)} (figure~\ref{fig:PixelTriplet}) and \emph{pixel quintuplets (pT5)} (figure~\ref{fig:PixelQuintuplet}), respectively.

After all of the objects have been created, a ``cross-cleaning" procedure is applied. This ensures that duplicates from different object collections are removed from the final output \emph{track candidate (TC)} collection and is implemented as follows: First, any pixel triplets that are close to any pixel quintuplet in the $\eta-\phi$ plane are marked as duplicates. The same procedure then takes place for quintuplets versus both pixel quintuplets and cleaned pixel triplets. Finally, remaining pixel line segments, unused in the previous linking steps, are cross-cleaned versus all the other objects. In the above cross-cleaning steps, it is also required that objects do not share any hits. As the last step of the algorithm, the track candidate collection is created by adding to it all of the following, non-duplicate objects:

\begin{itemize}
    \item pixel quintuplets,
    \item pixel triplets,
    \item quintuplets, and
    \item unlinked pixel line segments.
\end{itemize}

\subsection{Physics Performance} \label{sec:TheSegmentLinkingAlgo:PhysPerform}

\begin{figure}[tbh!]
\begin{center}
    \begin{minipage}{0.45\textwidth}
        \includegraphics[trim={0 0 3cm 3.6cm},clip,width=\textwidth]{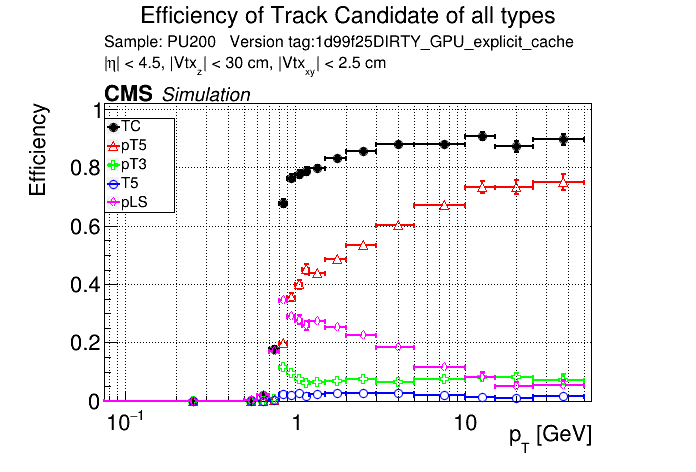}
        \caption{\label{fig:EffVsPt}Efficiency for reconstructed tracks in a \ttbar sample at 200 PU, as a function of the generator track \Pt. The color code is explained in the text.}
    \end{minipage}
    \hspace{2pc}
    \begin{minipage}{0.45\textwidth}
        \includegraphics[trim={0 0 3cm 3.6cm},clip,width=\textwidth]{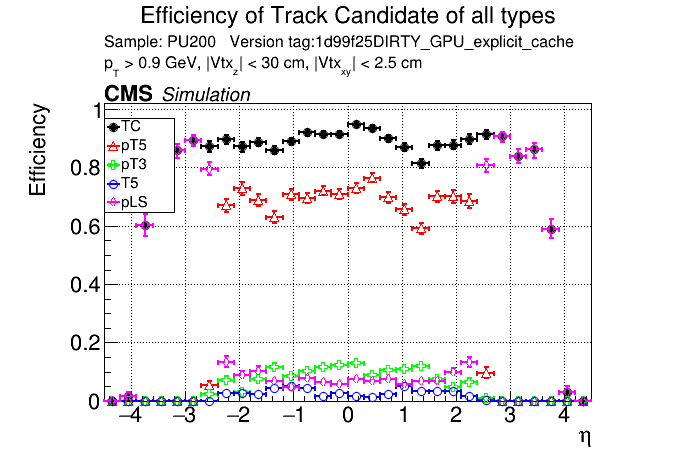}
        \caption{\label{fig:EffVsEta}Efficiency for reconstructed tracks in a \ttbar sample at 200 PU, as a function of the generator track $\eta$. The color code is explained in the text.}
    \end{minipage}
\end{center}
\end{figure}

\begin{figure}[tbh!]
    \begin{center}
        \includegraphics[trim={0 0 3cm 3.6cm},clip,width=0.45\textwidth]{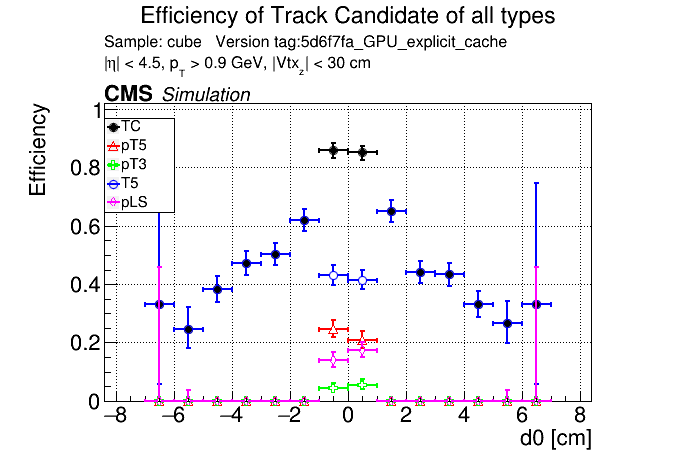}
        \hspace{2pc}
        \begin{minipage}[b]{0.45\textwidth}
            \caption{\label{fig:EffVsDxy}Efficiency for reconstructed tracks in a muon gun sample at 0 PU with the muon production vertex uniformly distributed in a cube with edge of $5~\text{cm}$, as a function of the generator track $d_{0}$, i.e. its distance from the center of the detector. The color code is explained in the text.}
        \end{minipage}
    \end{center}
\end{figure}

The selection of the objects that comprise the final track candidate collection is based on maximizing the track finding efficiency with the lowest possible fake rate. Each different object category aims at recovering efficiency of tracks with different characteristics. This is showcased in figures~\ref{fig:EffVsPt}-\ref{fig:EffVsDxy}, which show the breakdown of the track candidate efficiency (TC, in black) in different objects as a function of different variables.

Starting from figure~\ref{fig:EffVsPt}, which presents the efficiency as a function of \Pt, it is evident that the pixel quintuplets (pT5, in red), which are the longest objects, covering almost the whole range of the detector layers, are the driver of the performance. The same figure indicates that the pixel triplets (pT3, in green) and unlinked pixel line segments (pLS, in magenta) recover some efficiency at lower \Pt. This is because low \Pt\ tracks have a lower probability of leaving hits in a lot of tracker layers. The total efficiency for track candidates plateaus at its maximum value of $90\%$ already at $5 \GeV$. The real benefit of including the unlinked pixel line segments in the track candidate collection reveals itself in the efficiency plot as a function of $\eta$, where they noticeably are the drivers of the performance at the high $|\eta|$ range. Finally, figure~\ref{fig:EffVsDxy} shows that the inclusion of quintuplets (T5, in blue) is the unique handle of reconstructing displaced tracks.

\begin{figure}[tbh!]
\begin{center}
    \begin{minipage}{0.45\textwidth}
        \includegraphics[trim={0 0 3cm 3.6cm},clip,width=\textwidth]{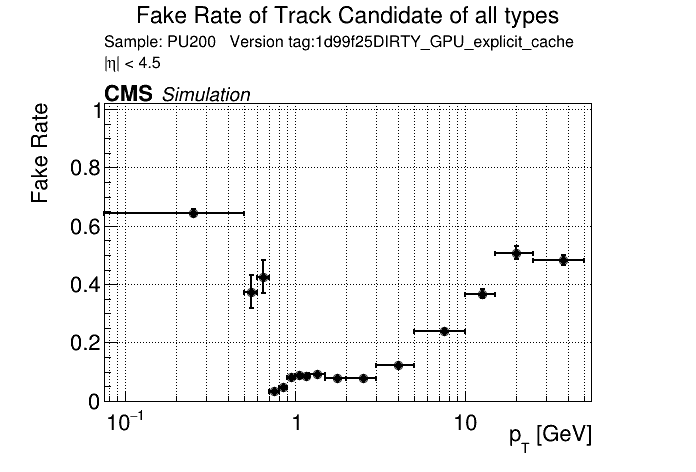}
        \caption{\label{fig:FRVsPt}Fake rate for reconstructed tracks in a \ttbar sample at 200 PU, as a function of the reconstructed track \Pt.}
    \end{minipage}
    \hspace{2pc}
    \begin{minipage}{0.45\textwidth}
        \includegraphics[trim={0 0 3cm 3.6cm},clip,width=\textwidth]{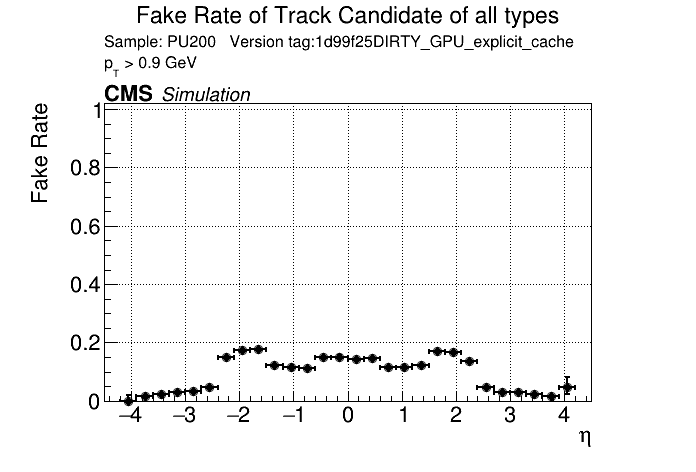}
        \caption{\label{fig:FRVsEta}Fake rate for reconstructed tracks in a \ttbar sample at 200 PU, as a function of the reconstructed track $\eta$.}
    \end{minipage}
\end{center}
\end{figure}

In terms of fake rate, figures~\ref{fig:FRVsPt} and \ref{fig:FRVsEta} present it as a function of \Pt\ and $\eta$ respectively. Above the lower \Pt\ threshold of $0.8 \GeV$, the fake rate remains less than $10\%$ for the bulk of the events, up till $\sim\!5 \GeV$. For higher \Pt\ values, some requirements on the objects are loosened to improve efficiency, which leads to an increase in the fake rate. The fake rate tends to be higher in the central region, where less accurate \Pt\ modules are employed with respect to the region of $|\eta| > 2.2$. Due to the extensive cross-cleaning, the duplicate rate is less than $5\%$, flat as a function of \Pt, with the main contribution coming from unused pixel segments at high $|\eta|$.

As a reference for the performance of the LST algorithm, the current CMS algorithms, adapted for HL-LHC, can be used. Looking at the ``Patatrack Trimmed" configuration of reference~\cite{CMSPhase2TrackingDP}, the efficiency of the LST algorithm is comparable with that, consistently around $90\%$ over the whole $|\eta|$ range. At the same time, the LST fake rate is higher, reaching $\sim\!15\%$ versus $4\%$ in the barrel and $\sim\!20\%$ versus $14\%$ in the endcap. It is worth mentioning that the comparison can only be very rough, since important components of track reconstruction, such as the track final fitting, have not been implemented yet for the LST algorithm.

\section{GPU Implementation} \label{sec:GPUImplementation}

GPUs have both advantages and disadvantages when compared to CPUs, making each one of these processing units suitable for different applications. GPUs host a much larger number of cores, $\mathcal{O}(10^3)$, than CPUs do, which have at most a few tens of cores. Due to this, GPUs can achieve greater numbers of floating point operations per second, hence being able to maintain high throughput. On the other hand, GPUs have limited memory resources, which can lead to high latency, when host-to-device data transfers are needed, while CPUs have lower memory latency, owing to more extensive use of caches. Given these characteristics, CPUs are more suitable for serial processing, while GPUs excel in parallel computations.

From the considerations above, it is clear that GPU programming requires both a parallelizable algorithm and an appropriate implementation to exploit the full capabilities of GPUs. As already mentioned, LST is an algorithm that is fit for parallelization according to the SIMD principle and our implementation, described below, aims to take advantage of the multitude of GPU cores while efficiently handling memory assignments and transfers. The LST algorithm implementation on GPU has been developed with CUDA, a programming framework for general-purpose computing on GPUs, developed by NVIDIA.

One cornerstone of the LST algorithm implementation is the usage of Structure of Arrays format, commonly referred to as ``SoA", with explicit memory management. Instead of using the common practice of object-oriented programming to create classes that hold the properties of a single object, separate arrays are created for separate object properties. These arrays hold each relevant quantity in continuous memory addresses, hence allowing for successive memory reads of the same quantity to be contiguous (memory coalescing). In this way, the time needed for accessing memory is minimized. Additionally, a custom cache memory allocation is utilized for small and commonly used objects. The cache memory is of very limited size but it allows for extremely fast memory transfers, helping further to decrease the total execution time of the algorithm. Finally, extensive studies on the final occupancy of all algorithm objects have led to truncations of their output numbers at the level of $99.9\%$ or $99.99\%$, resulting in a 5-fold reduction of the total memory footprint for the algorithm, while sacrificing only 1--$2\%$ of efficiency.

\begin{figure}[tbh!]
    \begin{center}
        \begin{tikzpicture}
        \matrix (m)[matrix of nodes, column  sep=-10mm,row  sep=8mm, align=center, nodes={rectangle,draw, anchor=center} ]{
            |[block]| {Hits} & & |[block]| {Stubs} & & |[block]| {Segments} \\
             & |[block]| {Triplets} & & |[block]| {Quintuplets} & \\
            |[block]| {Pixel Line Segments} & & |[block]| {Pixel Quintuplets} & & |[block]| {Pixel Triplets} \\
             & & |[block]| {Track Candidates} & & \\
        };
        \path [>=latex,->] (m-1-1) edge (m-1-3);
        \path [>=latex,->] (m-1-3) edge (m-1-5);
        \path [>=latex,->] (m-1-5) edge (m-2-2);
        \path [>=latex,->] (m-2-2) edge (m-2-4);
        \path [>=latex,->] (m-2-4) edge (m-3-1);
        \path [>=latex,->] (m-3-1) edge (m-3-3);
        \path [>=latex,->] (m-3-3) edge (m-3-5);
        \path [>=latex,->] (m-3-5) edge (m-4-3);
        \end{tikzpicture}
        \caption{\label{fig:KernelFlow}Flowchart of the order of kernel execution.}
    \end{center}
\end{figure}
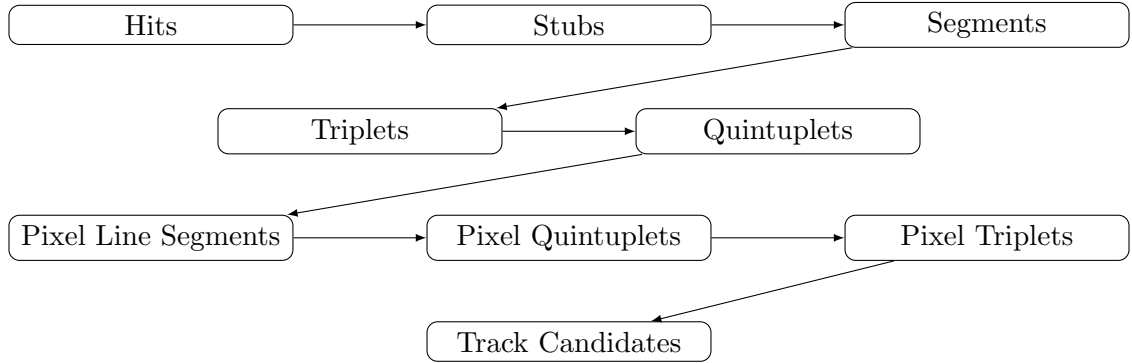

In terms of organization of its algorithmic steps, the LST code is split into different kernels for the creation of each object. A kernel is a set of operations that can run in parallel on independent GPU threads. Due to the fact that inputs for each kernel are highly localized, a large number of independent object reconstruction threads can run in the same kernel concurrently. The order in which kernels are launched roughly follows the description of the algorithm in subsection~\ref{sec:TheSegmentLinkingAlgo:AlgoLogic} and is shown in figure~\ref{fig:KernelFlow}.

\begin{figure}[tbh!]
    \begin{center}
        \includegraphics[width=\textwidth]{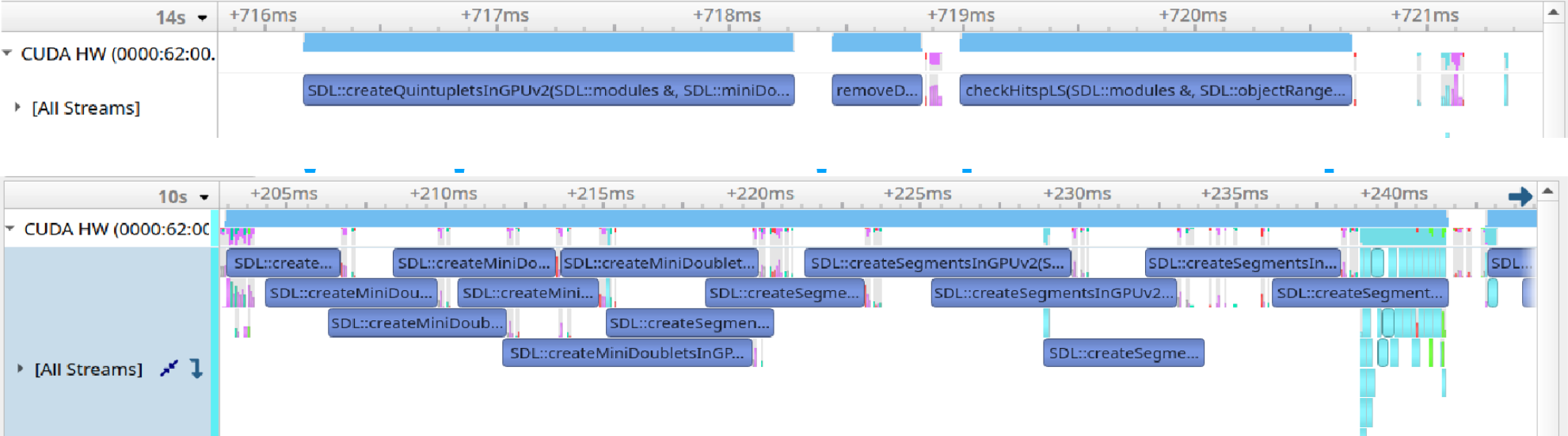}
        \caption{\label{fig:Multistreaming}Comparison of algorithm execution on 1-stream (upper) versus 8-streams (lower), using the NVIDIA Visual Profiler. Multi-streaming allows for running parts of multiple kernels concurrently.}
    \end{center}
\end{figure}

There is another layer of parallelization that can be exploited: event-level parallelization. Instead of running only one event in one GPU stream, multiple streams can be employed, each one of which runs a different event. This technique, called ``multi-streaming", allows for better utilization of the GPU resources and provides more opportunities for the kernel scheduler to insert more work. Moreover, it can hide the high GPU latency by assigning work to processing elements which already have data available, while others wait for data to be transferred. This can be seen in figure~\ref{fig:Multistreaming}, where light blue indicates the total work done by the GPU and dark blue shows the work done by kernels in different streams, both as a function of time. In the 1-stream case (upper schematic), the total GPU work is interrupted while the single kernel running waits for new data. On the other hand, in the 8-stream case (lower schematic), even though kernels in separate stream have similar waiting times, the total work done by the GPU is continuous in time, as data loading is hidden by work in other streams.

The benefits of multi-streaming are evident when measuring the time needed for the execution of the LST algorithm. This timing measurement is performed in a \ttbar sample at 200 PU and excludes the initial data transfer host-to-device for the hits, as well as any final fitting procedure on the output tracks, which has not been implemented in the current version of the algorithm. The average time needed for the execution of the algorithm on a single event is $32~\text{ms}$ for the 1-stream case, while it decreases to $26~\text{ms}$ for the 8-stream case, resulting in $\sim\!20\%$ improvement with respect to the 1-stream case. These timing measurements are in the same range as the latest CMS tracking timing measurements on CPUs~\cite{CMSPhase2TrackingDP}. It is worth noting that the LST algorithm is on par also price-wise with multi-CPU efforts, since two 32-core Intel Skylake Gold Xeon processors, commonly used in such efforts, cost approximately the same as one NVIDIA Tesla V100 GPU that has been used in our case.

\section{Summary and Outlook} \label{sec:SummaryAndOutlook}

In this work, Segment Linking (or Line Segment Tracking), a highly parallelizable track finding algorithm, aimed at alleviating the challenges that the HL-LHC will pose for tracking, has been presented. The algorithm has been successfully implemented on an NVIDIA Tesla V100 GPU and has been shown to have both physically and computationally comparable performance with cutting edge tracking efforts on CPUs.

More improvements of the algorithm are planned for the future. On the physics side, a re-evaluation of the selection criteria for objects is underway to improve the efficiency and lower the fake rate, especially for displaced tracks. Integration with other, similar GPU developments, such as Patatrack, is also foreseen. Computationally, mathematical optimizations for the calculation of physics parameters, as well as the extension of the usage of alternative data types, e.g. half-precision floats, are being explored. Plans to refine memory coalescing will potentially help reducing timing even further. Ultimately, the grand design for the algorithm involves its integration to the central CMS software for HLT and offline usage during the HL-LHC operation.

\section{Acknowledgements} \label{sec:Acknowledgements}

This work was supported by the U.S. National Science Foundation under Cooperative Agreements OAC-1836650 and PHY-2121686 and grant NSF-PHY-1912813.

\section*{References}
\bibliographystyle{iopart-num}
\bibliography{SegmentLinking_HEP2022Proceedings}

\end{document}